%% file: main.tex
\documentclass{article}
\usepackage{spconf,amsmath,amsfonts,graphicx,hyperref}
\usepackage{enumitem}
\usepackage{booktabs}
\usepackage{subcaption}
\usepackage{microtype}

\title{
  Neural Audio Codec for Robust Audio Deepfake Detection
}
\name{
  Jungwoo Kim$^{1,2}$\quad
  Joonyong Park$^{2,3}$\quad
  Junyoung Koh$^{1,2,4}$\quad
  Jong-Seok Lee$^{1}\footnotemark[1]$\sthanks{Corresponding author.}
}
\address{
  \begin{tabular}{c}
    $^{1}$Yonsei University, Seoul, Republic of Korea \\
    $^{2}$MAAP Lab, Republic of Korea \\
    $^{3}$The University of Tokyo, Tokyo, Japan \\
    $^{4}$University of Michigan, Ann Arbor, MI, USA \\
    \texttt{\small \{kjungwoo, jong-seok.lee\}@yonsei.ac.kr}
  \end{tabular}
}

\begin{document}
\ninept
\maketitle

\begin{abstract}
Audio deepfake detectors are typically evaluated on uncompressed audio, although real-world audio often undergoes low-bitrate coding.
In this work, we investigate \textbf{how audio coding affects deepfake detection} across codecs, bitrates, and detectors, finding higher errors at lower rates.
A mixed-pair protocol isolates codec-induced changes in bona fide and spoof audio, revealing asymmetric, codec-dependent failures: low-rate DAC and EnCodec mainly degrade bona fide detection, whereas X-Codec shows a stronger spoof-side limitation.
Motivated by these, we propose a \textbf{forensic-preserving neural audio codec (FP-NAC)}, which fine-tunes a pretrained codec using a detector-guided objective while preserving its native hard quantization path and bitrate.
On ASVspoof 2019 LA, FP-NAC reduces EER by up to 49.8~pp compared with the original DAC at 0.5~kbps while maintaining comparable reconstruction quality.
Although supervised by only one detector, FP-NAC improves performance across multiple detectors, highlighting forensic transparency as a codec design objective alongside perceptual quality.
Our codes are available at \url{https://github.com/kjungwoo03/FP-NAC}.
\end{abstract}


\input{sec/01_intro.tex}
\vspace{-1.0em}
\input{sec/02_rework.tex}
\vspace{-0.5em}
\input{sec/03_analysis.tex}
\input{sec/04_method.tex}
\input{sec/05_experiments.tex}
\input{sec/06_conclusion.tex}


\newpage
\clearpage

\bibliographystyle{IEEEbib}
\bibliography{refs}

\end{document}

%% file: sec/01_intro.tex
\section{Introduction}
\label{sec:intro}

The growing realism of synthetic speech has heightened concerns about deepfake misuse and motivated extensive research on audio deepfake detection~\cite{jung2022aasist, tak2021rawnet2}.
In practice, however, detectors often receive audio only after compression, transmission, or storage, rather than as a clean waveform.
More broadly, physical AI and autonomous systems increasingly process continuous multimodal sensory streams for both human and machine consumption~\cite{duan2020video, gao2021recent}.
Since machine-oriented coding can prioritize task-relevant information over exact signal fidelity, such settings further motivate low-bitrate and scalable coding under constrained bandwidth, latency, storage, and energy budgets~\cite{zeghidour2021soundstream, kim2025progressive, kim2026progressive}.
Neural audio codecs (NACs) achieve high perceptual quality at very low bitrates~\cite{zeghidour2021soundstream, defossez2023high, ye2025codec, kumar2023high}, but perceptual optimization does not necessarily preserve the forensic cues used for spoof detection.

Recent studies have begun to reveal interactions between NACs and audio deepfake detection.
Since modern speech generation systems often rely on NAC tokens or decoders~\cite{wu2024codecfake}, codec artifacts can appear in both bona fide and synthetic audio, potentially confounding codec-specific signatures~\cite{moussa2024unmasking} with spoofing cues~\cite{xiao2026dual}.
Conversely, codec token spaces have also been shown to contain information useful for identifying spoofed audio~\cite{li2024safeear, park2026probing, wu2026quantizer}.
These findings suggest that NAC representations are entangled with forensic cues rather than serving solely as perceptual compression representations.
Robustness studies have further shown that audio compression can substantially degrade detection performance, with neural codecs often causing severe performance drops even when perceptual quality remains high~\cite{li2026radd, li2024cross, shi2025benchmarking}.
These observations suggest that neural audio coding is not forensically transparent, as it can introduce codec-specific signatures while simultaneously altering spoof-detection evidence.

\begin{figure*}[t]
    \centering
    \vspace{-2.0em}
    \begin{subfigure}[t]{0.32\textwidth}
        \centering
        \includegraphics[width=\textwidth]{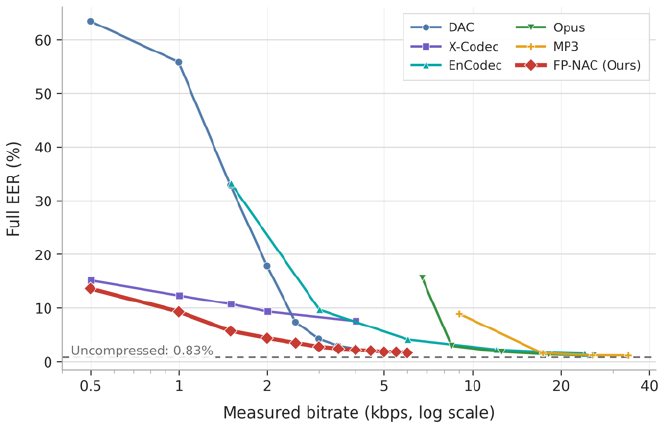}
        \label{fig:bitrate_aasist}
    \end{subfigure}
    \begin{subfigure}[t]{0.32\textwidth}
        \centering
        \includegraphics[width=\textwidth]{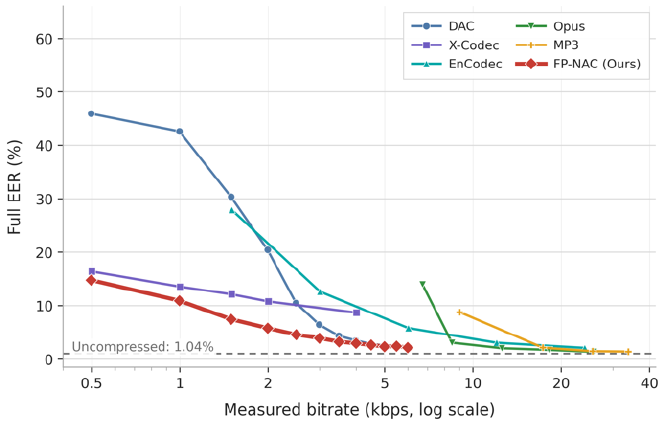}
        \label{fig:bitrate_aasistl}
    \end{subfigure}
        \vspace{-1.0em}
    \begin{subfigure}[t]{0.32\textwidth}
        \centering
        \includegraphics[width=\textwidth]{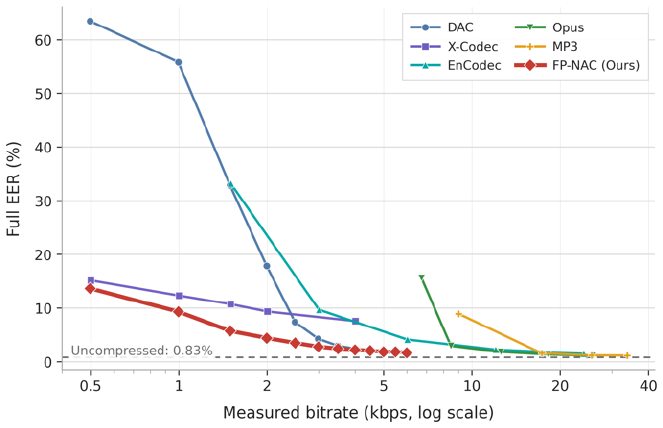}
        \label{fig:bitrate_rawnet2}
    \end{subfigure}
    \vspace{-0.75em}
    \caption{
        \textbf{Low-bitrate Coding Degrades Audio Deepfake Detection.}
        Full EER (\%) on the ASVspoof 2019 LA evaluation set~\cite{wang2020asvspoof} is reported with its measured bitrate across three different detectors. 
        The gray horizontal lines indicate the EER on uncompressed audio. 
        EER generally increases as bitrate decreases, while FP-NAC substantially reduces the EER compared with other NACs.
    }
    \label{fig:bitrate_eer}
    \vspace{-2.0em}
\end{figure*}

However, existing studies largely characterize this interaction through aggregate detection performance, leaving it unclear how coding alters forensic evidence---whether it primarily shifts bona fide speech toward spoof-like representations, suppresses spoof-specific evidence, or affects both sides simultaneously.
Existing mitigation is also predominantly detector-centric, requiring robustness to coding artifacts to be learned separately for each detector.
This raises a fundamental question: \textbf{How does audio coding distort the forensic evidence underlying deepfake detection, and can such evidence be preserved through codec design?}
We first introduce a mixed-pair evaluation protocol that separately probes codec-induced changes on bona fide and spoof trials.
Our analysis reveals strongly asymmetric and codec-dependent failure patterns: among the codecs studied, acoustic codecs fail predominantly on bona fide trials, whereas the semantically oriented X-Codec exhibits a comparatively stronger spoof-side failure.
To this end, we propose a \textbf{forensic-preserving neural audio codec (FP-NAC)}, a detector-guided codec adaptation that substantially reduces low-bitrate detection error and maintains comparable reconstruction quality, with gains that transfer to held-out detectors.

\begin{figure}[t]
    \centering
    \includegraphics[width=0.45\textwidth]{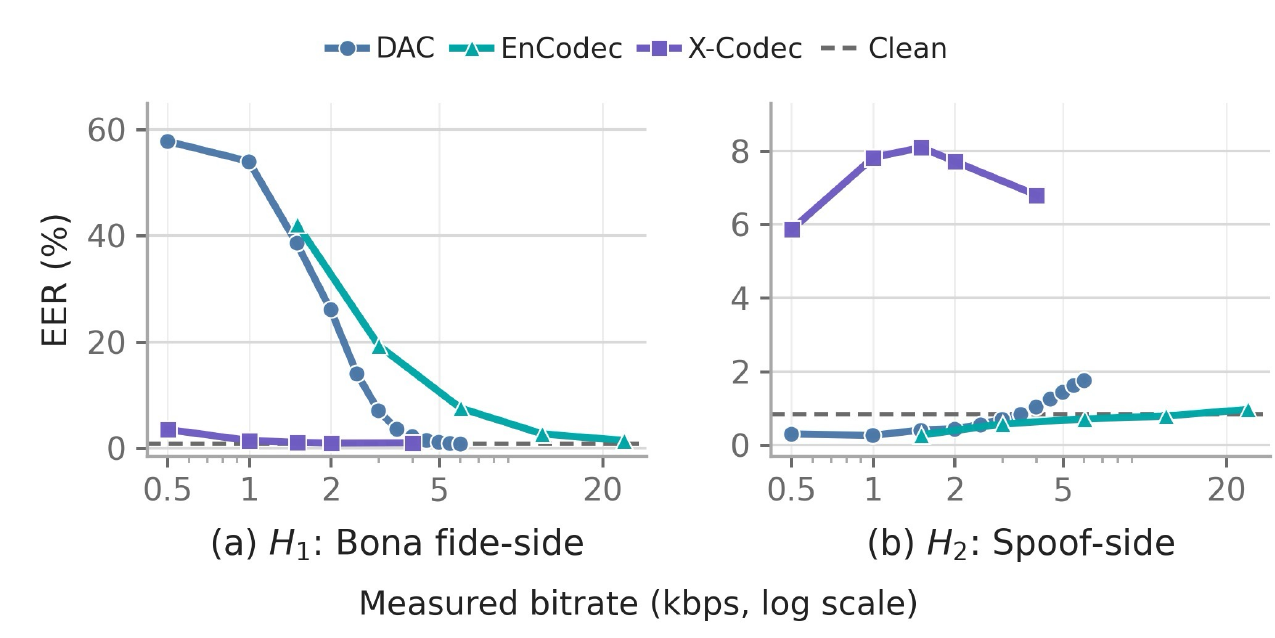}
    \vspace{-0.75em}
    \caption{
        \textbf{Codec-Dependent Asymmetric Failure Modes.} $H_1$ and $H_2$ across bitrates under AASIST~\cite{jung2022aasist}. 
        Acoustic codecs (DAC, EnCodec) show large $H_1$ that shrinks with bitrate, while X-Codec shows a persistently elevated $H_2$.
    }
    \label{fig:h1h2_bitrate}
    \vspace{-1.5em}
\end{figure}

Our contributions can be summarized as follows:
\begin{itemize}[leftmargin=8pt, topsep=0.8pt, itemsep=1pt, parsep=0pt]
    \item We introduce a mixed-pair evaluation protocol for diagnosing codec-induced forensic failures and show that different neural codecs exhibit distinct, strongly asymmetric class (bona fide vs. spoof)-side failure patterns.
    \item We propose FP-NAC, a parameter-efficient adaptation framework that explicitly incorporates forensic preservation into neural audio coding while retaining the pretrained codec backbone.
    \item We conduct extensive evaluations and mechanistic analyses, showing substantial low-bitrate EER reductions and consistent transfer to held-out detectors while maintaining reconstruction quality.
\end{itemize}

%% file: sec/02_rework.tex
\section{Related Work}
\label{sec:rework}

\textbf{Neural Audio Codecs.}
Neural audio codecs (NACs) compress waveforms into discrete representations through learned encoder--quantizer--decoder architectures.
SoundStream~\cite{zeghidour2021soundstream} and EnCodec~\cite{defossez2023high} employ residual vector quantization for scalable low-rate coding, while DAC~\cite{kumar2023high} emphasizes high-fidelity reconstruction and X-Codec~\cite{ye2025codec} incorporates semantic information.
However, these codecs are optimized primarily for perceptual quality rather than the preservation of task-specific forensic evidence.

\vspace{0.5em}
\noindent \textbf{Audio Deepfake Detection.}
Advances in speech generation, including neural codec language models, have increased the realism of synthetic speech and motivated audio deepfake detection~\cite{borsos2023audiolm, chen2025valle}.
Detectors such as AASIST~\cite{jung2022aasist} and RawNet2~\cite{tak2021rawnet2} use complementary representations but remain vulnerable to unseen attacks and domain shifts~\cite{li2024cross}.
Recent studies show that codec artifacts and spoof-relevant information can severely degrade detection~\cite{wu2024codecfake, xiao2026dual, li2026radd}. 

%% file: sec/03_analysis.tex
\section{Mixed-Pair Evaluation of Codec-Induced Forensic Failures}
\label{sec:analysis}

We first examine how audio coding affects deepfake detection across different codecs and bitrates.
As shown in Fig.~\ref{fig:bitrate_eer}, detection performance degrades substantially as bitrate decreases.
This trend is consistently observed across multiple detectors, suggesting that codec-induced degradation is not specific to a particular detector design.
However, the conventional Full EER only measures the separability between coded bona fide and coded spoof samples, and does not reveal which class-side evidence is primarily affected by coding.

\vspace{0.5em}
\noindent\textbf{Protocol Design.}
Let $\mathcal{B}$ and $\mathcal{S}$ denote clean bona fide and spoof samples, respectively, and $\mathcal{C}(\cdot)$ denote codec processing.
In addition to Full EER between $\mathcal{C}(\mathcal{B})$ and $\mathcal{C}(\mathcal{S})$, we define $H_1$: $\mathrm{EER}(\mathcal{C}(\mathcal{B}), \mathcal{S})$, which pairs coded bona fide audio with clean spoofed audio to isolate how strongly coding pushes bona fide evidence toward the spoof side, and $H_2$: $\mathrm{EER}(\mathcal{B}, \mathcal{C}(\mathcal{S}))$, which pairs clean bona fide audio with coded spoofed audio to isolate how strongly coding suppresses spoof-specific cues.
These are independently thresholded diagnostic EERs and are not additive components of Full EER.

\begin{figure}[t]
    \centering
    \vspace{0.5em}
    \includegraphics[width=0.45\textwidth]{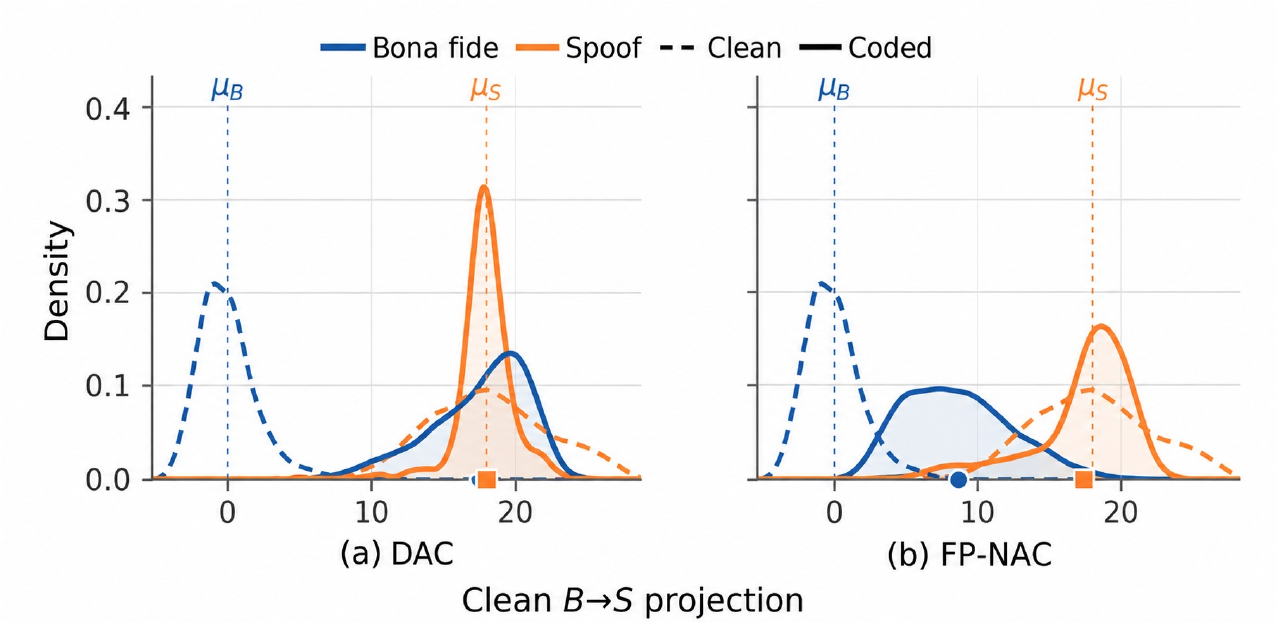}
    \vspace{-1.0em}
    \caption{
    \textbf{Class-Directed Representation Shifts in AASIST.}
    Detector embeddings at 0.5~kbps are projected onto the clean bona fide--spoof centroid axis.
    DAC strongly shifts bona fide representations toward spoof, while FP-NAC substantially mitigates this shift.
    }
    \label{fig:embedding_shift}
    \vspace{-1.5em}
\end{figure}

\begin{figure*}[t]
    \centering
    \vspace{-0.75em}
    \includegraphics[width=0.85\linewidth]{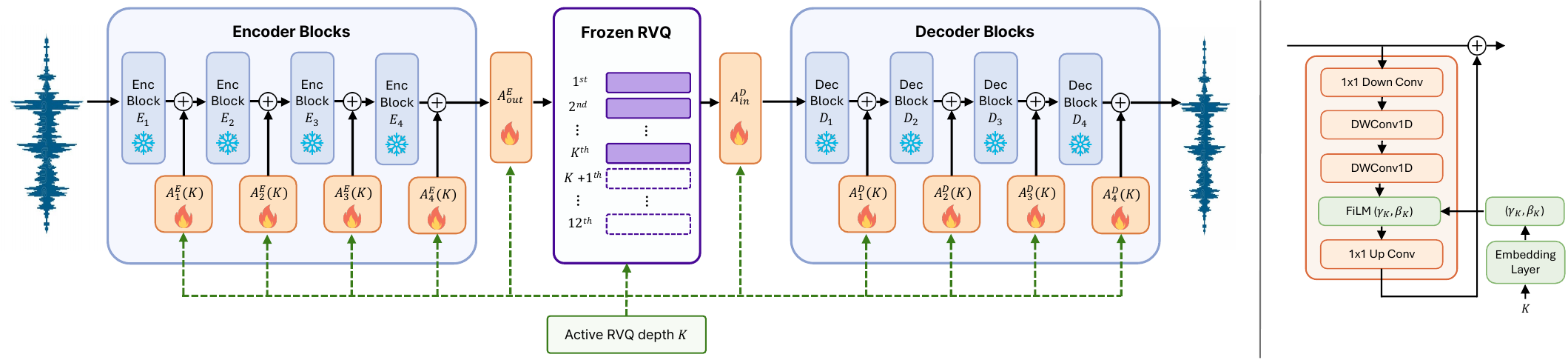}
    \vspace{-0.8em}
    \caption{
    \textbf{Overall Architecture of FP-NAC.}
    The left panel shows frozen DAC backbones with trainable residual adapters conditioned on the active RVQ depth $K$.
    The right panel illustrates the adapter architecture, consisting of bottleneck projection, depthwise temporal convolutions, $K$-conditioned FiLM modulation, and residual addition.
    }
    \label{fig:architecture}
    \vspace{-1.5em}
\end{figure*}

\vspace{0.5em}
\noindent\textbf{Asymmetric Failure Modes.}
Fig.~\ref{fig:h1h2_bitrate} shows that low-bitrate DAC~\cite{kumar2023high} and EnCodec~\cite{defossez2023high} exhibit high $H_1$ but low $H_2$.
This is consistent with codec-specific artifacts acting as spoof-like signatures~\cite{xiao2026dual}: they can make coded bona fide audio more spoof-like (high $H_1$) while reinforcing the spoof decision for coded spoof audio (low $H_2$).
Hence, a low $H_2$ does not necessarily imply that the codec preserves the spoof-like evidence.
In contrast, X-Codec~\cite{ye2025codec} shows low $H_1$ but high $H_2$, where coding introduces less spoof-like distortion to bona fide audio while suppressing the spoof-discriminative evidence.
These contrasting trends suggest that codec-induced forensic degradation is not merely a consequence of bitrate reduction, but also depends on how different codec designs transform class-discriminative evidence.

\vspace{0.5em}
\noindent\textbf{Detector-Space Interpretation.}
To examine how this asymmetry appears inside the detector, we project the 160-dimensional AASIST~\cite{jung2022aasist} embeddings onto the axis directly connecting the clean bona fide and spoof centroids.
Unlike a generic dimensionality reduction (\textit{e.g.}, PCA), this centroid-defined axis provides an interpretable clean class-separation direction, where movement toward the spoof centroid corresponds to a spoof-directed representation shift.
As shown in Fig.~\ref{fig:embedding_shift}(a), DAC moves the bona fide distribution close to the clean spoof region, while coded spoof representations exhibit little movement toward bona fide along this axis.
Consequently, coded bona fide and clean spoof become highly overlapping ($H_1$), whereas clean bona fide and coded spoof remain well separated ($H_2$), explaining the $H_1$-dominant failure in Fig.~\ref{fig:h1h2_bitrate}.

Overall, these results show that codec-induced forensic degradation is class-asymmetric, rate-dependent, and directionally structured in detector space.
This motivates a codec-side intervention that preserves detector-relevant evidence for both bona fide and spoofed audio across operating rates.

%% file: sec/04_method.tex
\section{Proposed Method}
\label{sec:method}

In this section, we propose a \textbf{forensic-preserving neural audio codec (FP-NAC)}, illustrated in Fig.~\ref{fig:architecture}.
Motivated by the class-asymmetric and rate-dependent failures identified above, FP-NAC aims to preserve detector-relevant evidence for both bona fide and spoofed audio.
It freezes the DAC backbone and inserts $K$-conditioned residual adapters on both sides of the RVQ, enabling rate-specific corrections while retaining the original coding path.

\subsection{Rate-Conditioned Residual Adaptation}

As shown in the left panel of Fig.~\ref{fig:architecture}, encoder-side adapters are inserted after each encoder block and immediately before the residual vector quantization (RVQ), allowing adaptation of the representations presented to the quantizer.
Decoder-side adapters are placed immediately after the RVQ and after each decoder block, modifying how the selected discrete codes are realized in the reconstructed waveform.

Given an intermediate feature $h\in\mathbb{R}^{C\times T}$, each adapter projects $h$ into a compact bottleneck and applies depthwise temporal processing, as illustrated in the right panel of Fig.~\ref{fig:architecture}.
Here, $C$ denotes the number of feature channels and $T$ denotes the temporal length.
To account for different quantization regimes across bitrates, we condition each adapter on the active RVQ depth $K$ using feature-wise linear modulation (FiLM)~\cite{perez2018film}:
\begin{equation}
    [\gamma_K,\beta_K] = W_{\text{FiLM}} e_K+b_{\text{FiLM}},
\end{equation}
\begin{equation}
    A_K(h)
    =
    h+\alpha_K W_{\mathrm{up}}
    \left(
    (1+\gamma_K)\odot G(W_{\mathrm{down}}h)+\beta_K
    \right),
\end{equation}
where $W_{\mathrm{FiLM}}$, $W_{\mathrm{up}}$, and $W_{\mathrm{down}}$ are learnable weight matrices, $b_{\mathrm{FiLM}}$ is a learned bias term, $e_K$ is a learned embedding of $K$, $G(\cdot)$ denotes the depthwise temporal processing blocks, and $\alpha_K$ is a learnable rate-specific residual gate.
We initialize $\alpha_K$ as 0, such that $A_K(h)=h$ at initialization.
Conditioning on $K$ allows a single checkpoint to learn rate-dependent corrections across different operating bitrates.

\subsection{Forensic-Preserving Optimization}
\label{sec:optimization}

\noindent\textbf{Soft-Backward RVQ.}
Building on straight-through vector quantization and differentiable soft-to-hard relaxations~\cite{agustsson2017soft}, we retain hard code assignments in the forward pass while using a soft-assignment surrogate only to backpropagate gradients to the encoder-side adapters.
The surrogate is removed at inference, preserving hard RVQ assignments and leaving the original DAC bitstream format and nominal bitrate unchanged.

\vspace{0.5em}
\noindent\textbf{Forensic Preservation Objective.}
We jointly optimize reconstruction quality and forensic preservation:
\begin{equation}
    \mathcal{L}_{\mathrm{total}}
    =
    \mathcal{L}_{\mathrm{quality}}
    +
    \mathcal{L}_{\mathrm{forensic}}.
\end{equation}
$\mathcal{L}_{\mathrm{quality}}$ uses a DAC-style objective~\cite{kumar2023high} consisting of Mel reconstruction, waveform fidelity, RVQ commitment, adversarial, and discriminator feature-matching losses.
For forensic preservation, we employ a frozen pretrained detector $F$ and define
\begin{equation}
    \mathcal{L}_{\mathrm{forensic}}
    =
    r(t)\left(
    \lambda_{\mathrm{det}}\mathcal{L}_{\mathrm{det}}
    +
    \lambda_{\mathrm{feat}}\mathcal{L}_{\mathrm{feat}}
    +
    \lambda_{\mathrm{env}}\mathcal{L}_{\mathrm{env}}
    \right),
\end{equation}
where $r(t)$ gradually introduces forensic supervision during early training.

Let $s_F(x)$ denote the bona fide score of $F$, and let $\theta$ and $\rho$ denote the EER threshold and score scale estimated from clean development data.
With $\bar{s}_F(x)=(s_F(x)-\theta)/\rho$ and $y_i\in\{+1,-1\}$ for bona fide and spoof trials, respectively, we define
\begin{equation}
    \mathcal{L}_{\mathrm{det}}
    =
    \frac{1}{N}\sum_{i=1}^{N}
    \max\left(
    0,\,
    m-y_i\bar{s}_F(\hat{x}_i)
    \right),
\end{equation}
where $\hat{x}_i$ denotes the reconstructed waveform of the $i$-th input, $N$ is the minibatch size, and $m$ is the detection margin.
This finite-margin objective stops penalizing samples once the desired class margin is satisfied, avoiding unnecessary amplification of detector-specific scores.
In addition, $\mathcal{L}_{\mathrm{feat}}$ preserves detector representations between the original and reconstructed signals, while $\mathcal{L}_{\mathrm{env}}$ preserves the smooth spectral envelope of spoof trials.
The detector remains frozen throughout training and serves only as a forensic teacher.

\begin{figure}[t]
    \centering
    \vspace{-0.5em}
    \includegraphics[width=0.375\textwidth]{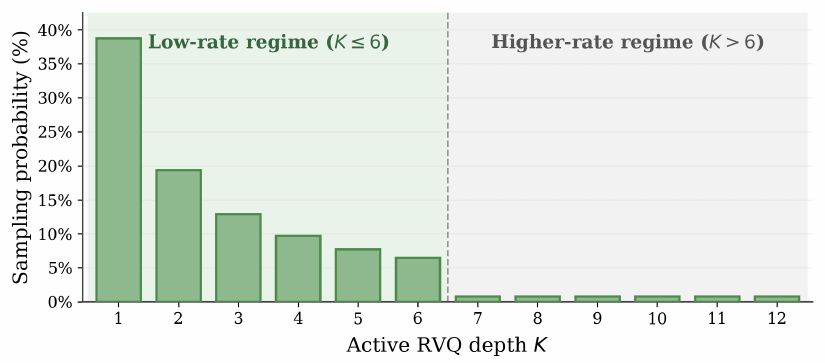}
    \vspace{-1.0em}
    \caption{
        \textbf{Multi-Rate Sampling Distribution.} Sampling probability over the active RVQ depth $K$.
    }
    \label{fig:rate_sampling}
    \vspace{-1.5em}
\end{figure}

\vspace{0.5em}
\noindent\textbf{Multi-Rate Training.}
To support multiple operating rates with a single checkpoint, FP-NAC is jointly optimized over $K=1,\ldots,12$ using the quantizer-dropout-style multi-rate training~\cite{zeghidour2021soundstream} with the rate-dependent sampling distribution shown in Fig.~\ref{fig:rate_sampling}.
Reflecting the stronger forensic degradation at lower rates, the distribution places greater training emphasis on smaller $K$ while retaining coverage of the full rate range.

%% file: sec/05_experiments.tex
\section{Experiments}
\label{sec:experiments}

\subsection{Experimental Setup}
\label{subsec:experimental-setup}

\noindent\textbf{Dataset and Detectors.}
We train and evaluate FP-NAC on ASVspoof 2019 LA~\cite{wang2020asvspoof}.
AASIST~\cite{jung2022aasist} is used as the frozen forensic teacher $F$ during training, while AASIST-L~\cite{jung2022aasist} and RawNet2~\cite{tak2021rawnet2} are held-out detectors for evaluating cross-detector transfer of forensic preservation.

\begin{figure}[t]
    \centering
    \includegraphics[width=0.35\textwidth]{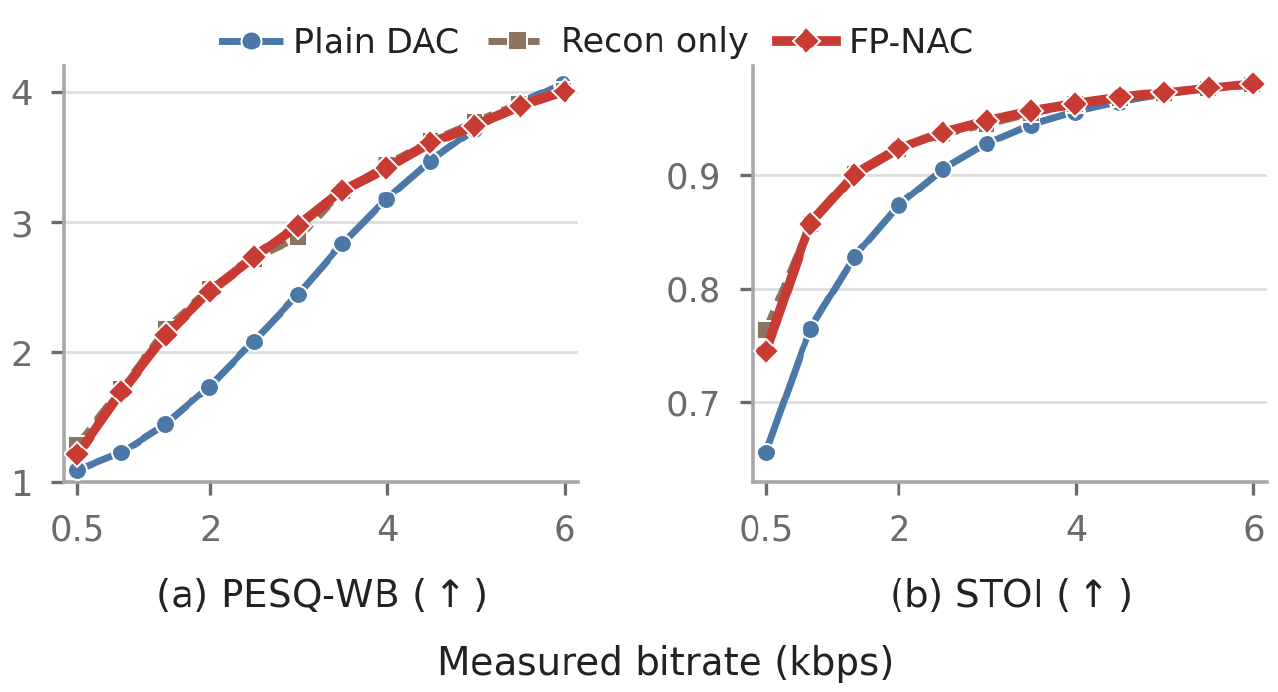}
    \vspace{-1.0em}
    \caption{
        \textbf{Reconstruction Quality across Bitrates.}
        PESQ-WB~\cite{itu_p862_2} and STOI~\cite{taal2011algorithm} on a subset of ASVspoof 2019 LA~\cite{wang2020asvspoof}.
       Both reconstruction-only adaptation and FP-NAC improve low-rate reconstruction quality over Plain DAC, with closely matched quality across most operating points.
    }
    \label{fig:recon_quality}
    \vspace{-1.5em}
\end{figure}

\vspace{0.5em}
\noindent\textbf{Codecs and Metrics.}
We compare FP-NAC with the original DAC~\cite{kumar2023high}, EnCodec~\cite{defossez2023high}, X-Codec~\cite{ye2025codec}, and conventional codecs, Opus~\cite{opuswebsite} and MP3~\cite{iso11172_3}.
For detection, each waveform is standardized to 64,600 samples (about 4.04~s at 16~kHz): longer utterances are truncated to the first 64,600 samples, while shorter ones are zero-padded.
Detection performance is measured by EER.

\vspace{0.5em}
\noindent\textbf{Training Details.}
FP-NAC is trained for 48 epochs using AdamW~\cite{loshchilov2019decoupled} with learning rates of $5\times10^{-5}$ and $5\times10^{-6}$ for the encoder- and decoder-side adapters, respectively.
We set 0.5 of $\lambda_{\mathrm{det}}$, 0.1 of $\lambda_{\mathrm{feat}}$, 0.05 of $\lambda_{\mathrm{env}}$, and the detection margin $m$ as 0.5.
The forensic objective is linearly warmed up during the first 8 epochs. 

\subsection{Main Results}
\label{sec:main_results}

\noindent\textbf{Low-Rate Detection Performance.}
As shown in Fig.~\ref{fig:bitrate_eer}, FP-NAC substantially reduces the severe detection degradation of the original DAC in the low-rate regime.
At 0.5~kbps, EER decreases from 63.47\% to 13.66\% on AASIST, from 45.93\% to 14.78\% on AASIST-L, and from 25.74\% to 13.20\% on RawNet2.
Importantly, although forensic supervision uses only AASIST, the low-rate gains consistently transfer to both held-out detectors.

\vspace{0.5em}
\noindent\textbf{Reconstruction Quality.}
We further evaluate reconstruction quality using PESQ-WB~\cite{itu_p862_2} and STOI~\cite{taal2011algorithm} on a fixed class-balanced subset of 1,024 evaluation utterances (512 bona fide and 512 spoof).
For comparison, we additionally train a \textit{reconstruction-only} variant, using the same adapter architecture and multi-rate training scheme as FP-NAC, but without the forensic objective $\mathcal{L}_{\mathrm{forensic}}$.

As shown in Fig.~\ref{fig:recon_quality}, both reconstruction-only and FP-NAC substantially improve low-to-mid-rate reconstruction quality over the original DAC, while their PESQ-WB and STOI remain closely matched across most operating points.
This indicates that FP-NAC achieves its forensic gains without a substantial reconstruction quality trade-off, while most of the quality improvement itself comes from reconstruction-oriented adaptation.
At higher rates, the differences diminish as all variants approach the original DAC performance.

\vspace{0.5em}
\noindent\textbf{Detector-Space Recovery.}
As shown in Fig.~\ref{fig:embedding_shift}(b), FP-NAC substantially moves the coded bona fide distribution away from the spoof region, while coded spoof representations remain on the spoof side.
This restores much of the codec-domain class separation collapsed by Plain DAC and is consistent with the large reduction in the $H_1$-dominant failure at low bitrates.

\begin{table}[t]
    \centering
    \caption{
    \textbf{Parameter Overhead of Adapters.}
    Additional trainable parameters relative to the 74.14M-parameter original DAC~\cite{kumar2023high}.
    }
    \label{tab:adapter_params}
    \vspace{-0.75em}
    \footnotesize
    \setlength{\tabcolsep}{4pt}
    \begin{tabular}{lcc}
    \toprule
    Configuration & Trainable Params. & Overhead \\
    \midrule
    Plain DAC     & 74.14M  & 0\% \\
    Encoder only  & +0.72M  & +0.97\% \\
    Decoder only  & +0.58M  & +0.78\% \\
    Both (FP-NAC) & +1.30M  & +1.75\% \\
    \bottomrule
    \end{tabular}
    \vspace{-1.0em}
\end{table}

\begin{figure}[t]
    \centering
    \includegraphics[width=0.4\textwidth]{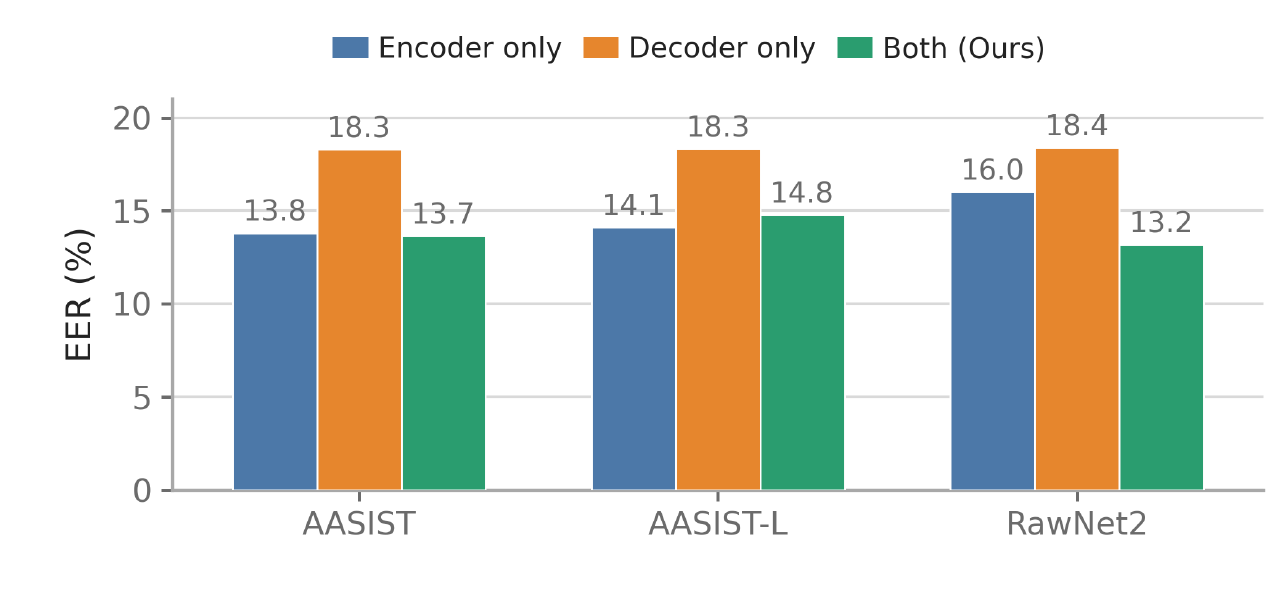}
    \vspace{-2.0em}
    \caption{
    \textbf{Encoder--Decoder Adaptation Ablation at 0.5~kbps.}
    Full EER (\%) across AASIST, AASIST-L, and RawNet2 for encoder-only, decoder-only, and joint adaptation (ours).
    }
    \label{fig:position_ablation}
    \vspace{-1.5em}
\end{figure}

\subsection{Ablation Studies}
\label{sec:ablation}

\noindent\textbf{Adaptation Target.}
Fig.~\ref{fig:position_ablation} compares encoder-only, decoder-only, and joint adaptation at 0.5~kbps.
Encoder-side adaptation alone provides substantial recovery, while joint adaptation achieves the best average performance across the three detectors, supporting adaptation on both sides of the RVQ.
As summarized in Table~\ref{tab:adapter_params}, FP-NAC introduces only 1.30M trainable parameters (1.75\% of the original DAC), compared with 0.72M (0.97\%) and 0.58M (0.78\%) for encoder-only and decoder-only adaptation, respectively.

\vspace{0.5em}
\noindent\textbf{Multi-Rate Sampling.}
Replacing the low-rate-focused sampling distribution in Fig.~\ref{fig:rate_sampling} with uniform sampling improves most operating points at $K \geq 4$, reducing the higher-rate average EER from 3.24\% to 2.91\%.
However, uniform sampling substantially degrades performance at $K=1$, increasing the EER by 5.62~pp.
Mixed-pair evaluation reveals a trade-off: uniform sampling lowers $H_2$, but its higher $H_1$ at low rates indicates greater bona fide collapse.
Despite comparable average PESQ-WB, the two sampling strategies emphasize different rate regimes: uniform sampling favors medium and higher rates, whereas the proposed sampling prioritizes the $H_1$-dominant failure at $K=1$.
Given that codec-induced detection degradation is substantially more severe at low bitrates, the proposed distribution prioritizes the regime in which forensic preservation is more needed.

\vspace{0.5em}
\noindent\textbf{$\mathbf{\textit{K}}$-Conditioning.}
To isolate the effect of explicit rate conditioning, we remove the $K$-dependent FiLM layer while keeping all other settings unchanged.
Removing $K$ conditioning leaves low-rate performance nearly unchanged, but increases the higher-rate average EER from 3.24\% to 3.87\% and reduces PESQ-WB from 3.65 to 2.93.
Because the proposed sampling distribution emphasizes low-$K$ samples, removing explicit $K$ information causes low-rate-dominated updates to impose a shared correction across all rates.
Conditioning instead enables rate-specific corrections, reducing cross-rate interference and limiting degradation at larger $K$.

%% file: sec/06_conclusion.tex
\section{Conclusion}
\label{sec:conclusion}

In this work, we systematically investigated how low-bitrate audio coding affects deepfake detection, revealing asymmetric and codec-dependent failure modes through mixed-pair evaluation.
Motivated by this, we proposed FP-NAC, which adapts the original DAC using detector-guided forensic supervision.
On ASVspoof 2019 LA, FP-NAC reduced EER by up to 49.8~pp at 0.5~kbps, transferred consistently to held-out detectors, and maintained comparable objective reconstruction quality.